\documentclass[aps,twocolumn,10pt,superscriptaddress,prl,floatfix]{revtex4-2}

\usepackage{bm}
\usepackage{graphicx}
\usepackage{subfig}
\usepackage{natbib}
\usepackage{siunitx}
\usepackage{mathtools}
\usepackage{bbold} 
\usepackage{color}
\usepackage{dsfont}

\newcommand{\ket}[1]{\left| #1 \right>}

\begin{document}

\title{Phase sensitive quantum spectroscopy with high frequency resolution}

\author{Nicolas Staudenmaier}
\email{nicolas.staudenmaier@uni-ulm.de}
\affiliation{Institute for Quantum Optics, Ulm University, D-89081 Ulm, Germany}
\author{Simon Schmitt}
\affiliation{Institute for Quantum Optics, Ulm University, D-89081 Ulm, Germany}
\author{Liam P. McGuinness}
\email{liam.mcguinness@anu.edu.au}
\affiliation{Institute for Quantum Optics, Ulm University, D-89081 Ulm, Germany}
\affiliation{Laser Physics Centre, Research School of Physics, Australian National University, Acton, Australian Capital Territory 2601, Australia}
\author{Fedor Jelezko}
\affiliation{Institute for Quantum Optics, Ulm University, D-89081 Ulm, Germany}
\affiliation{Center for Integrated Quantum Science and Technology, Ulm University, D-89081 Ulm, Germany}

\begin{abstract}
Classical sensors for spectrum analysis are widely used but lack micro- or nanoscale spatial resolution. On the other hand, quantum sensors, capable of working with nanoscale precision, do not provide precise frequency resolution over a wide range of frequencies. Using a single spin in diamond, we present a measurement protocol for quantum probes which enables full signal reconstruction on a nanoscale spatial resolution up to potentially 100\,GHz. We achieve $58\,\mathrm{nT/\sqrt{Hz}}$ amplitude and $0.095\,\mathrm{rad/\sqrt{Hz}}$ phase sensitivity and a relative frequency uncertainty of $10^{-12}$ for a 1.51\,GHz signal within 10\,s of integration. This technique opens the way to quantum spectrum analysis methods with potential applications in electron spin detection and nanocircuitry in quantum technologies. 
\end{abstract}

\maketitle

Spectrum analysis, whereby phase, amplitude or frequency information is extracted from periodic signals, is a widespread tool underpinning applications ranging from imaging and microscopy \cite{Betzig1992}, chemical identification \cite{Aue1976, Wuethrich2001, Spiess2008}, development of time and frequency standards \cite{Diddams2004} quantum state tomography \cite{Lloyd2014}, radar detection \cite{Turpin1981} and medical diagnosis \cite{Lauterbur1973, Bowtell2008}. As a physical measurement is required to provide information to spectral estimation algorithms, their ultimate accuracy is governed by physical laws with limits given by quantum mechanics. Detectors made up of individual atoms thereby allow information encoded in such spectra to be obtained at the limits of sensitivity, resolution and non-invasiveness. 

Here, we construct a protocol which allows a single quantum coherent spin to form a heterodyne detector (quantum analogue of the classical heterodyne detector) for  near-resonant fields. We extend techniques recently developed to improve the spectral resolution at low frequencies $<$\,100\,MHz based on dynamical decoupling \cite{Schmitt2017, Boss2017, Glenn2018}, to high frequencies in the microwave regime where existing methods for microwave detection are limited in terms of spatial or spectral resolution \cite{Chipaux2015, Horsley2016, Horsley2018}. We use single spins associated with nitrogen-vacancy (NV) centers in diamond to perform spectroscopy of magnetic fields oscillating at gigahertz frequencies, close to the spin resonance frequency. At the nanoscale, single NV centers have allowed for nuclear magnetic resonance and electron paramagnetic resonance spectroscopy of single molecules and nuclei to be performed \cite{Grinolds2013, Mamin2013, Staudacher2013}. Importantly, by recording the frequency of the magnetic field, structural and spatial information of the sample can be obtained \cite{Mamin2013, Staudacher2013, Aue1976, Wuethrich2001}. Applications of the presented technique could be for detection of single electron spins \cite{Grinolds2013, Hall2016}, spin waves of magnons \cite{Finco2021}, characterization of miniaturized electric circuits for communication and quantum technologies \cite{Bardin2021} and (Doppler) radar detection \cite{Alabaster2012}.

While the analogy to classical heterodyne detection is not perfect, the protocol preserves many of the same hallmarks. Namely:
\begin{enumerate}
\item down-conversion of high frequency signals to bandwidth within the readout bandwidth,
\item simultaneous recording of phase, amplitude and frequency information, allowing for complete signal reconstruction,
\item frequency resolution limited by the stability of an external clock, detector independent.
\end{enumerate}
We demonstrate each of these characteristics by constructing an atomic heterodyne detector from a single quantum coherent spin. 

The idea is to tailor the sensor-signal interaction in such a way that the result of each measurement depends on the phase of the signal. This is done by introducing a local oscillator which can be used to obtain a beat-note with the signal. We refer to the technique as high frequency Qdyne due to the analogy to classical heterodyne detection but with a quantum sensor \cite{Schmitt2017}. Note that this technique is different to recent methods requiring dynamical decoupling as here no qubit control is performed during sensing duration \cite{Pang2017, Schmitt2017, Boss2017, Meinel2021}. Methods for low frequency detection rely on dynamical decoupling methods where the sensor accumulates a phase that is transferred into different populations. In our approach instead the signal directly drives the sensor transition. Hence, there is no need to employ dynamical decoupling. Analogues to another technique \cite{Chu2021, Meinel2021} can be drawn, although here we use a distinct protocol which is immediately applicable to sensing of continuous fields. In the present work we focus on the phase resolving capabilities and characterize the protocol's performance for spectrum analysis. 

\begin{figure*}[tbh]
\begin{center}
\includegraphics[width=0.95 \textwidth]{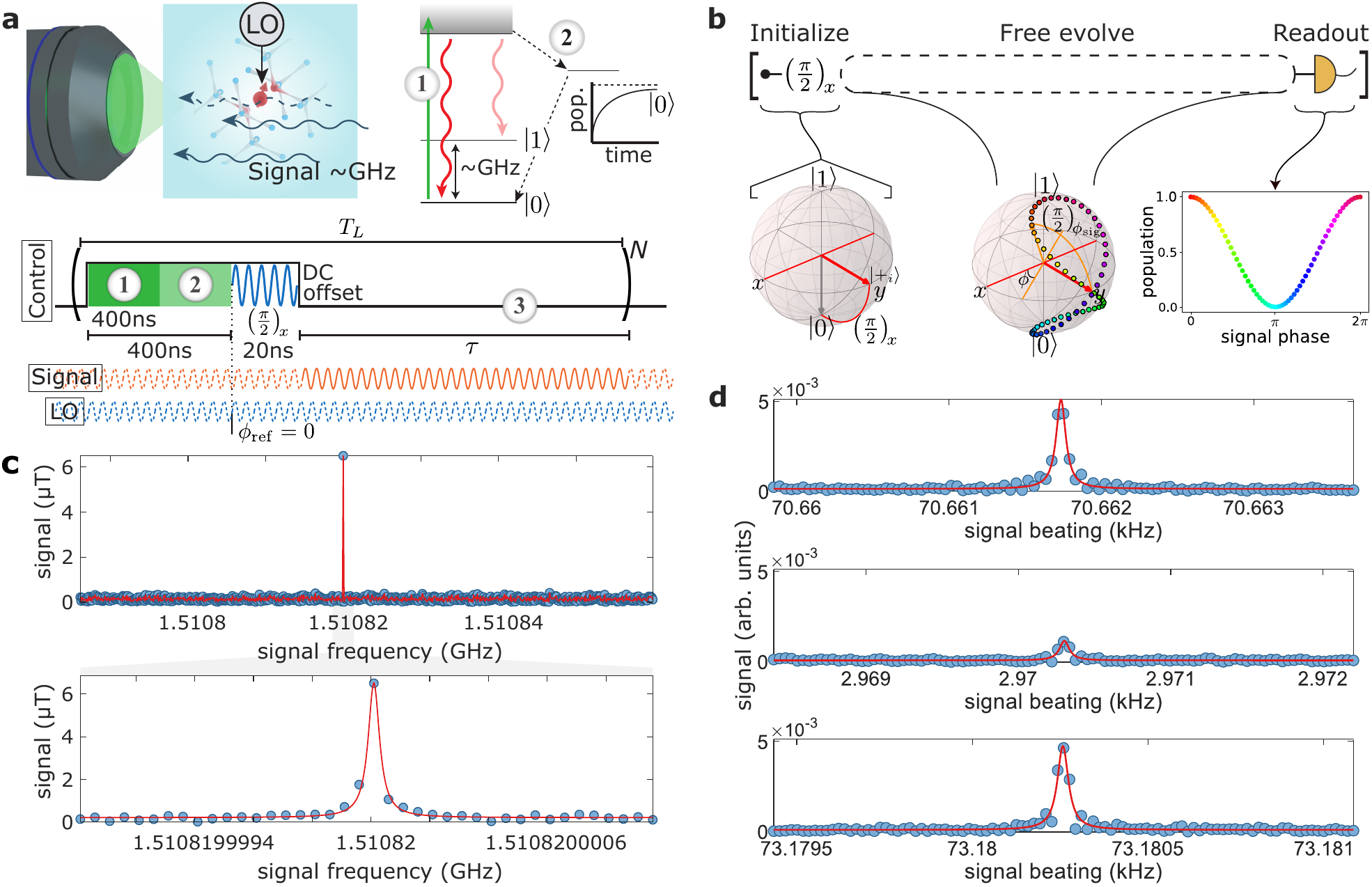}
\protect\caption{\small Nanoscale high frequency sensing. \textbf{(a)} Confocal setting for sensing with a single NV center in diamond. The right panel shows its level structure and the lower panel the pulse sequence for high frequency Qdyne. (1) The NV center is excited with a green laser and the fluorescence is collected. (2) Population decay via a metastable state initializes the NV center in $\ket{0}$. (3) A DC shift can be applied to drive the sensor out of resonance with respect to the signal. The offset is turned off for a time $\tau$ to allow interaction with the sensor. The single sequence length of duration $T_L$ is repeated many times. \textbf{(b)} Working principle of the high frequency Qdyne method. After preparation of $\ket{+_i}$ with a $\pi/2$-pulse around the $x$-axis the final state after signal interaction depends on the signal phase $\phi$. A signal interaction resulting in a $\pi/2$ rotation is shown. The states are shown on the Bloch sphere in the rotating frame of the local oscillator. \textbf{(c)} Measurement of a 1.51082\,GHz signal. In a sampling time of ten seconds three million samples are taken that correspond to one and a half million frequency channels in the FFT (only every 2000th point is shown). Lower plot: zoom around the peak. \textbf{(d)} Measurement with a signal being applied only during sensing period (top) and having a continuous signal without (center) and with DC shift (bottom). The $x$-axis is the absolute value $\delta = |\delta_0|$ of the beat-note of Eq.\,\eqref{eq:beating}.}
\label{fig1}
\end{center}
\end{figure*}

\subsection{1) High frequency Qdyne technique}
The magnetic field of a near-resonant signal is described with a time varying function of $B(t) = B_0 \cos(2\pi\nu_\mathrm{sig}t + \phi_0)$ with unknown amplitude $B_0$, phase $\phi_0$ and frequency $\nu_\mathrm{sig}$. In the rotating frame the interaction Hamiltonian with a two-level system is written as 
\begin{equation}
H = \frac{\hbar \Delta}{2} \, \sigma_z + \frac{\hbar\Omega_\mathrm{sig}}{2} \, \sigma_{\phi _0}
\label{eq:Hamiltonian}
\end{equation}
where $\sigma_{\phi_0} = \cos\phi_0 \, \sigma_x + \sin\phi_0 \, \sigma_y$, the Pauli matrices $\sigma_i$, the driving amplitude $\Omega_\mathrm{sig} = \sqrt{\Delta^2 + \Omega_0^2}$ with detuning $\Delta = 2\pi (\nu_\mathrm{sig} - \nu_\mathrm{sens})$, the frequency difference between the signal and the sensor's resonance $\nu_\mathrm{sens}$, and $\Omega_0 = \gamma_\mathrm{sens}B_0$ (gyromagnetic ratio $\gamma_\mathrm{sens}$ of the sensor). 

To sense this signal we use a single nitrogen-vacancy (NV) center in diamond. Its spin ground state can be effectively described as a two-level system with states $\ket{0}$ and $\ket{1}$. Readout is done optically with a green laser and collecting the spin-dependent fluorescence (see Figure \ref{fig1}a). After initialization in $\ket{0}$, the NV center is prepared in the superposition state $\ket{+_i} = \frac{1}{\sqrt{2}} \, (\ket{0} + i\ket{1})$ by a $\pi/2$-pulse around the $x-$axis by a reference pulse with known phase. This state then evolves under the action of the near-resonant signal field for some time $\tau$. The protocol is constructed such that the reference pulse always has the phase $\phi_\mathrm{ref} = 0$ and for every subsequent repetition the same state $\ket{+_i}$ is prepared. We note that this is not a stringent requirement for the protocol, since it can also be achieved with a signal generator where $\phi_\mathrm{ref}$ changes according to the control frequency, but as presented here it is mathematically streamlined, and experimentally achievable using an arbitrary waveform generator. 

In the following, if not otherwise stated, we toggle the signal on during the $\tau$ interval, and off for the remaining time (as performed in Ref. \cite{Meinel2021}) with a switch. However, for many spectroscopy applications one may not have control over the signal. To circumvent signal deterioration for continuous fields (especially for strong signals) one can apply a DC magnetic field that shifts the sensor transition out of resonance during state preparation and readout. 

Assuming signal interaction only occurs during $\tau$ the population in $\ket{1}$ of the final state is calculated to be 
\begin{equation}
\begin{aligned}
|c_1|^2 = \frac{1}{2} \bigg [ &1 - 
	\frac{\Delta\Omega_0}{\Omega_\mathrm{sig}^2} (1 - \cos(\Omega_\mathrm{sig}\tau)) \sin(\phi) + \\
&\frac{\Omega_0}{\Omega_\mathrm{sig}} \sin(\Omega_\mathrm{sig}\tau) \cos(\phi) \bigg ]
\label{eq:spin_population}
\end{aligned}
\end{equation}
where $\phi = \phi(t) = 2\pi\nu_\mathrm{sig}t + \phi_0$ is an instantaneous phase of the signal at some time $t$. For calculation of this result see Supplemental Material at [\textit{URL will be inserted by publisher}]. Assuming small detuning, $\Delta \ll \Omega_0$, Eq.\,\eqref{eq:spin_population} is approximated as
\begin{equation}
|c_1|^2 \approx \frac{1}{2} [1 + \sin(\Omega_0 \tau) \cos(\phi)].
\label{eq:spin_population_approx}
\end{equation} 

Experimentally, the spin population is sampled at fixed time intervals $T_L$ such that the signal phase $\phi$ changes by a constant increment and each outcome is stored individually. It is this synchronization of single measurements at a rate of $1/T_L$, in addition to the $\phi_\mathrm{ref}$, that defines the local oscillator (LO) frequency:
\begin{equation}
\nu_\mathrm{LO} = \frac{\mathrm{round}(\nu_\mathrm{sens} T_L)}{T_L} = \frac{N_\mathrm{LO}}{T_L}. 
\label{eq:LO}
\end{equation}
$N_\mathrm{LO} = \mathrm{round}(\nu_\mathrm{sens} T_L)$ is the rounded integer number of periods of the sensor resonance frequency $\nu_\mathrm{sens}$ within $T_L$ and the number that defines the closest local oscillator frequency to a signal frequency that is within the linewidth of the sensor. A detailed discussion of the local oscillator can be found in the Supplemental Material. We emphasize that the local oscillator could also be defined by a control field that is used to manipulate the NV center at resonance with a continuously updating phase \cite{Chu2021, Meinel2021} but then a reference phase $\phi_\mathrm{ref} \neq 0$ has to be considered for each measurement. 

Sampling at times $T_n = n\cdot T_L$, the populations $|c_{1,n}|^2$ are calculated from Eq.\,\eqref{eq:spin_population} (or \eqref{eq:spin_population_approx}) with phases $\phi_n = \phi_0 + 2\pi\,T_n \delta_0$ where the phase increment is determined by the beating of the signal against the local oscillator 
\begin{equation}
\delta_0 = \nu_\mathrm{sig} - \nu_\mathrm{LO}.
\label{eq:beating}
\end{equation}
As a result, the outcome probability of each measurement oscillates with frequency $\delta_0$. 

To show the working principle, a diamond sample fabricated into a solid immersion lens that is overgrown with a 100\,nm thick layer of isotopically purified \textsuperscript{12}C is used. The high purity allows the NV center to reach dephasing times up to 50\,$\mu$s. An external magnetic field is aligned along the NV-axis to about 50\,mT. The sample is mounted in a confocal microscope setup that is controlled via the software suite \textit{Qudi} \cite{Qudi}. In Figure \ref{fig1}c a 1.51\,GHz signal (approximately NV resonance frequency) is measured and a Fast Fourier Transform (FFT) resolves the oscillation in the sampled data. After preparing $\ket{+_i}$ the signal of 6.5\,$\mu$T amplitude interacts with the NV center for 1404\,ns and in an integration time of ten seconds 3$\times$10\textsuperscript{6} samples are obtained. From the local oscillator the frequency channels of the FFT can be assigned to the scanned spectrum. 

In Figure \ref{fig1}d, we compare the measured spectrum when the signal is toggled on and off, as opposed to applied continuously. In the upper panel, the signal is only on during $\tau$, and in the middle panel, the signal remains on continuously. In the lower panel, the signal is applied continuously, but the NV center is shifted out of resonance by application of a DC shift which is only turned off during $\tau$. The DC shift is created by applying a constant current to the NV center control stripline which causes a magnetic field of one Gauss at the NV center and shifts the resonance frequency by 3\,MHz. As the control $\pi/2$-pulse is much stronger than the sensing field, only the signal is brought out of resonance by the DC shift. We find that a continuously applied signal but with DC control gives comparable results to one toggled on/off, while it significantly reduces the measurement signal without DC control.

\subsection{2) Signal reconstruction}
\begin{figure}[t]
\begin{center}
\includegraphics[width=0.48 \textwidth]{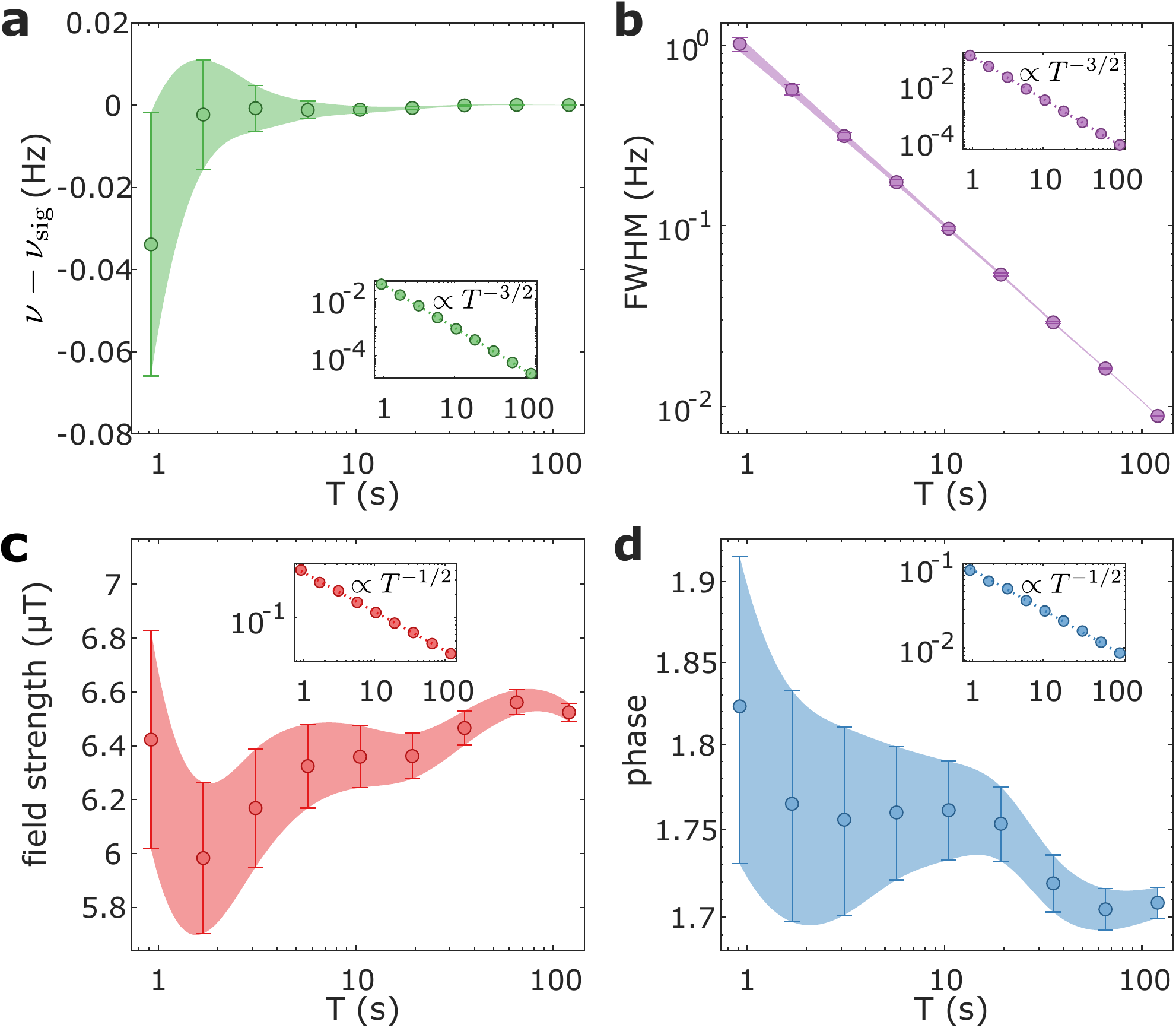}
\protect\caption{\small Full signal reconstruction. \textbf{(a), (c), (d)} signal frequency, strength and phase estimation. Measuring all three parameters allows full reconstruction of the signal. In (a) the difference between the estimated frequency $\nu$ and the signal frequency $\nu_\mathrm{sig}$ is displayed. \textbf{(b)} Linewidth (full width at half maximum) $\Delta\nu = 1/T$. The insets show the uncertainty of the respective estimation (error bars) as the 95\,\% confidence intervals of the fit parameters. (a, b) scale with $T^{-3/2}$ and (c, d) with $T^{-1/2}$.} 
\label{fig2}
\end{center}
\end{figure}

Analysis of the discrete Fourier transform, just as in classical heterodyne detection, allows for full signal reconstruction. In Figure \ref{fig2} estimation of the signal parameters after different integration times is shown for the same signal as shown in Figure \ref{fig1}c. Frequency and amplitude (a, c) are estimated with a Lorentzian fit on the peak in the FFT. The uncertainty is obtained from the 95\,\% confidence interval of the fit. Importantly, frequency uncertainty scales as $T^{-3/2}$ as a product of reduced noise ($T^{-1/2}$) and reduced linewidth ($1/T$) as a function of total measurement time $T$. A precision $<1\,\mathrm{mHz}$ is achieved after 10\,s integration time, resulting in a relative frequency uncertainty $\frac{\delta\nu}{\nu} < 10^{-12}$. Ultimately, the resolution and uncertainty are limited by the stability of the clock which times the arbitrary waveform generator, i.e. the local oscillator defined via the sequence length produced by the arbitrary waveform generator. As the timing is given to an accuracy $10^{-7}$ the estimated signal frequency may be systematically shifted from the actual frequency. As a result the analysis only includes statistical errors and not systematic ones. 

For phase estimation due to the finite resolution in the FFT spectrum, the signal phase is estimated from the phases at the two frequency channels next to the estimated peak (obtained from the argument of the complex valued FFT). Then, with linear interpolation the phase of the beat-note is obtained, from which the initial signal phase $\phi_0$ can be recovered because the local oscillator's reference is $\phi_\mathrm{ref} = 0$. Note that for a detuned signal the estimated phase can be different from the signal phase $\phi_0$. From Eq.\,\eqref{eq:spin_population} one can see that the beat-note is the sum of a sine and cosine. The resulting oscillation has the same frequency but not necessarily the same phase $\phi$. In our measurements the signal detuning $\Delta$ is small and this effect can be neglected (see Eq.\,\eqref{eq:spin_population_approx}). In the Supplemental Material [\textit{URL will be inserted by publisher}] a detailed analysis of this phase shift is included. 

Attention has to be given to the sign of $\delta_0$ in Eq.\,\eqref{eq:beating}. If $\delta_0 < 0$ a negative frequency is sampled. Thus, the signal phase is the negative of the estimated one from the FFT. Furthermore, frequency estimation might not be unique because it does not distinguish between negative and positive frequency. The observed signal has the frequency $\delta = |\delta_0| = |\nu_\mathrm{sig} - \nu_\mathrm{LO}|$. This ambiguity can be resolved with a second measurement as outlined in the Supplemental Material.

\subsection{3) Spectrum analysis benchmarks}
\subsection{}
The measurement technique allows for spectrum analysis on the nanoscale as it employs an atomic sized sensor that can be used to estimate signals within a wide frequency range. To understand the capabilities that arise with the technique we investigate some benchmarks of this protocol. 

\begin{figure}
\begin{center}
\includegraphics[width=0.48\textwidth]{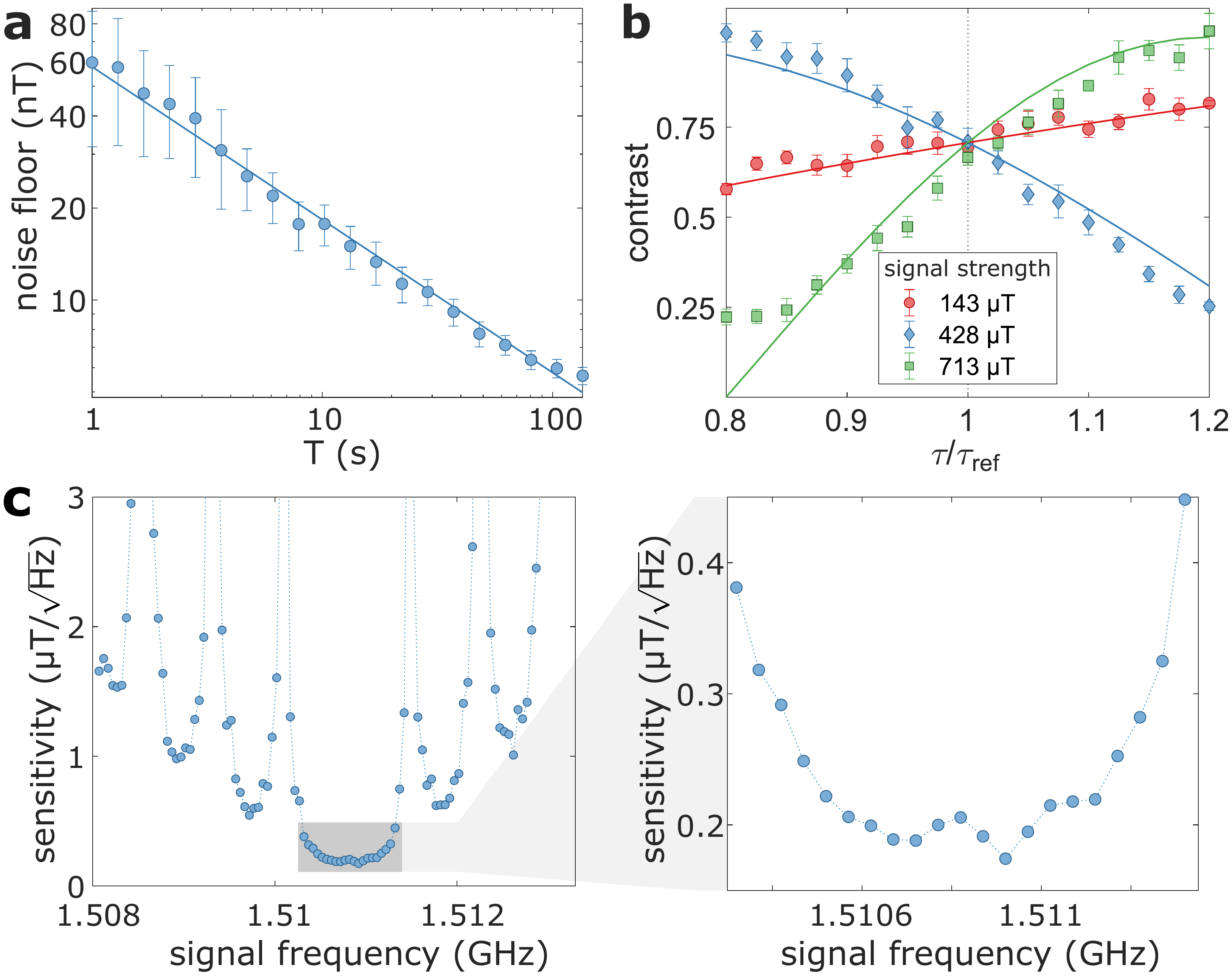}
\protect\caption{\small \textbf{(a)} Noise floor in dependence on integration time. Amplitude sensitivity of the measurement is obtained from a fit of slope $T^{-1/2}$ that gives $58\,\mathrm{nT}$ noise floor at $T=1\,\mathrm{s}$. \textbf{(b)} Example of signal strength ambiguities for an interaction length $\tau_\mathrm{ref} = 31.3\,\mathrm{ns}$. These ambiguities are lifted when the interaction time is changed. \textbf{(c)} Sensitivity in dependence on the signal frequency. Signals of strength $6.5\,\mathrm{\mu T}$ with varying frequency are detected. All other measurement parameters are kept constant.}
\label{fig3}
\end{center}
\end{figure}

Standard quantum sensing protocols usually rely on the pointwise accumulation of spectral information \cite{Degen2017}. In contrast, this technique makes intrinsic use of FFT-mode with a sample rate $1/T_L$. In this way, information over many frequency channels is sampled over the measurement time. This allows the sensitivity to be increased by minimizing dead-time. In Figure \ref{fig3}a the noise floor in the FFT spectrum of a measurement with 223\,nT signal strength is shown from which a sensitivity of $58\,\mathrm{nT/\sqrt{Hz}}$ is obtained (noise floor at $T=1\,\mathrm{s}$). From the data of Figure \ref{fig2} we calculate a sensitivity of $0.095\,\mathrm{rad/\sqrt{Hz}}$ for phase and $0.03\,\mathrm{Hz/Hz^{3/2}}$ for frequency estimation. We want to emphasize that no dynamical decoupling is incorporated here. High frequency pulsed \cite{Joas2017} or continuous decoupling \cite{Stark2017} offer the potential to further increase the amplitude sensitivity to 4\,$\mathrm{nT/\sqrt{Hz}}$ \cite{Balasubramanian2009}. 

We further investigate the spectral and dynamic range and bandwidth of the technique. The operating range can be extended from megahertz (or even less) to high frequency ($>$\,GHz) by tuning the resonance of the sensor. For the NV center a range from 1\,GHz to 5\,GHz can be covered with static magnetic fields up to 70\,mT and 100\,GHz are reached with a 3.5\,T field \cite{Stepanov2015, Aslam2015}. The upper limit to the operating range is set by technical challenges that are involved for generating stable high magnetic fields and handling high frequency microwaves, the lower limit is set simply by the integration time $1/T$. 

The dynamic range is set by the sensor properties and integration time. An upper limit can be clearly defined because each interaction with a signal which results in a rotation of more than $\pi/2$ of the sensor state cannot be distinguished from a rotation less than $\pi/2$. For that reason we set the upper end of the dynamic range to the signal strength at which a $\pi/2$ rotation of the sensor within $\tau = T_2^* / 2$ (the sensing time for which best sensitivity is obtained \cite{Degen2017}) is performed, $B_\mathrm{max} = \frac{\pi}{\gamma_\mathrm{NV} T_2^*}$. With $\gamma_\mathrm{NV} = 2\pi \times 28.03\,\mathrm{MHz/mT}$ and $T_2^*=50\,\mathrm{\mu s}$ we have $B_\mathrm{max} = 0.36\,\mathrm{\mu T}$. However, one always has the freedom to increase the dynamic range by setting $\tau < T_2^* / 2$ at the expense of a reduced sensitivity as is done in the measurement of Figure 1c. By using multiple measurements with different $\tau$ it is also possible to increase the dynamic range because ambiguities for a rotation larger and smaller than $\pi/2$ get resolved, and with a lower trade-off in reduced sensitivity. This is measured for three different signal strengths in Figure \ref{fig3}b that result in the same measurement signal for an interaction time of $\tau_\mathrm{ref} = 31.3\,\mathrm{ns}$. The lower end of the dynamic range is simply given by the sensitivity, since this is the minimum field strength that can be identified. The full dynamic range for a single measurement is now $58\,\mathrm{nT} \times \sqrt{T/1\,\mathrm{s}}$ (total integration time $T$) to $360\,\mathrm{nT}$ for $\tau = T_2^* / 2 = 25\,\mathrm{\mu s}$. 

The bandwidth is naturally given by the linewidth of the sensor's transition and is limited by the dephasing time $\Delta\nu_b = 1/T_2^*$. However, the bandwidth might be reduced if temporal overheads are involved for preparation or readout of the sensor. In these cases a unique assignment of the frequency within the sensor's linewidth might not be possible, due to undersampling. If the sequence length exceeds the dephasing time, $T_L = \tau + t_\mathrm{overhead} > T_2^*$, the sampling rate limits the effective bandwidth $\Delta\nu_b = 1/T_L$. Note that we distinguish between positive and negative $\delta_0$ and that the sampling theorem $\delta_\mathrm{max} = \delta_{0,\mathrm{max}} = -\delta_{0,\mathrm{min}} = 1 / 2T_L$ is always satisfied. Furthermore, for a very weak signal, the condition $\Omega_0 > \Delta$ may not be satisfied, although the signal frequency lies within the detector bandwidth. Thus the signal strength can also define the bandwidth as $\Omega_0 / 2\pi$. 

In figure \ref{fig3}c the sensitivity is measured in dependence of the signal frequency while all other measurement parameters are kept constant. The protocol is susceptible also to frequencies outside of the sensor linewidth but with a reduced sensitivity. In the Supplemental Material we present a method to overcome potential ambiguities. 

In comparison to classical spectrum analyzers this quantum analogue yields some specific peculiarities. While classical devices can have better sensitivity they are much larger in size and operate far from nanoscale spatial resolution. As an atomic sized defect in the diamond lattice, the NV center (or other appropriate quantum sensors) are able to measure small signals on the nanoscale with nanometer spatial resolution. In order to be sensitive to a certain frequency bandwidth, an external magnetic field has to be adjusted correctly. Classical frequency swept analyzers can cover a large bandwidth by applying band-limited electrical filters. The drawback in this case is that the analyzer may require a long time to record the spectrum step by step. FFT mode analyzers on the other side yield a bandwidth that is given by the sampling rate. Our quantum spectrum analyzer also operates in FFT mode and the bandwidth is given by either the sensor linewidth or the sampling rate. Applying this Qdyne technique to the work of Chipaux et al. \cite{Chipaux2015} where a magnetic field gradient is used with an NV center ensemble, a wide frequency range can be covered with the drawback of reducing nanoscale to microscale resolution.

\subsection{Discussion}
We have presented a powerful measurement technique that extends high frequency resolution quantum sensing to high frequency oscillating fields. In analogy to classical heterodyne detection full signal reconstruction is possible as frequency, amplitude and phase information is provided. We obtain a phase sensitivity of $0.095\,\mathrm{rad/\sqrt{Hz}}$. Detailed analysis of the spectral and dynamic range show possibilities and limitations of spectrum analysis for different signals and can be adapted to various probes used as quantum sensor. Detection of highly coherent signals will benefit from the technique especially in settings on the nanoscale such as miniaturized integrated circuits for communication or quantum technologies. Velocimetry in Doppler Radar detection for velocities down to a few $\mu\mathrm{m/s}$ \cite{Jing2020} in nano- to microscale settings will be possible thanks to the high spectral resolution and phase sensitivity. Methods to increase spectral and dynamic range are presented that allow to overcome some of their limitations with help of a second measurement. Employing NV center ensembles with a high magnetic field gradient could be used to considerably increase the bandwidth of the method.

\begin{acknowledgments}
\subsection{Acknowledgment}
We acknowledge fruitful discussions with Genko T. Genov. 

During the preparation of the manuscript we became aware of the related work by J. Meinel et al. \cite{Meinel2021}. 

This work is supported by the Australian Research Council, Future Fellowship (FT180100100) (L.P.M.), the Bosch Forschungsstiftung (N.S. and F.J.), the European Research Council via Synergy Grant HyperQ, the German Research Foundation (excellence cluster POLIS and CRC1279), the German Federal Ministry of Education and Research and the European Commission via ASTERIQS (F.J.).
\end{acknowledgments}

\bibliographystyle{unsrt}

\clearpage
\onecolumngrid

\begin{center}
{\Large \textbf{Supplemental Material}}
\end{center}

\section{Experimental Setup}
\subsection{Diamond sample}
For the experiments a hemispherical diamond prepared by Element Six is used which was overgrown with a thin layer ($\sim$100\,nm) of 99.999\,\% isotopically purified \textsuperscript{12}C. The NV centers are incorporated during the growth process. With the high purity of \textsuperscript{12}C atoms a dephasing time of about $T_2^*=50\,\mathrm{\mu s}$ is obtained. 

\subsection{Optical and microwave setup}
The centerpiece of the optical setup is a home-built confocal microscope with diffraction limited resolution. A 532\,nm laser is used to excite the single NV centers and the red-shifted fluorescence is detected by an avalanche photodiode. Microwave control is generated with an arbitrary waveform generator (Tektronix AWG70001A) with 20\,ps timing resolution. Magnetic signal fields at frequencies $\sim$\,GHz are generated using a crystal quartz oscillator stabilized signal generator (Rohde \& Schwarz SMIQ~03B). All signals are combined and guided through a 20\,\textmu m thin copper wire soldered on the sample holder and located close to the NV centers of interest. 

\begin{figure}[t]
\includegraphics[width = 0.8\textwidth]{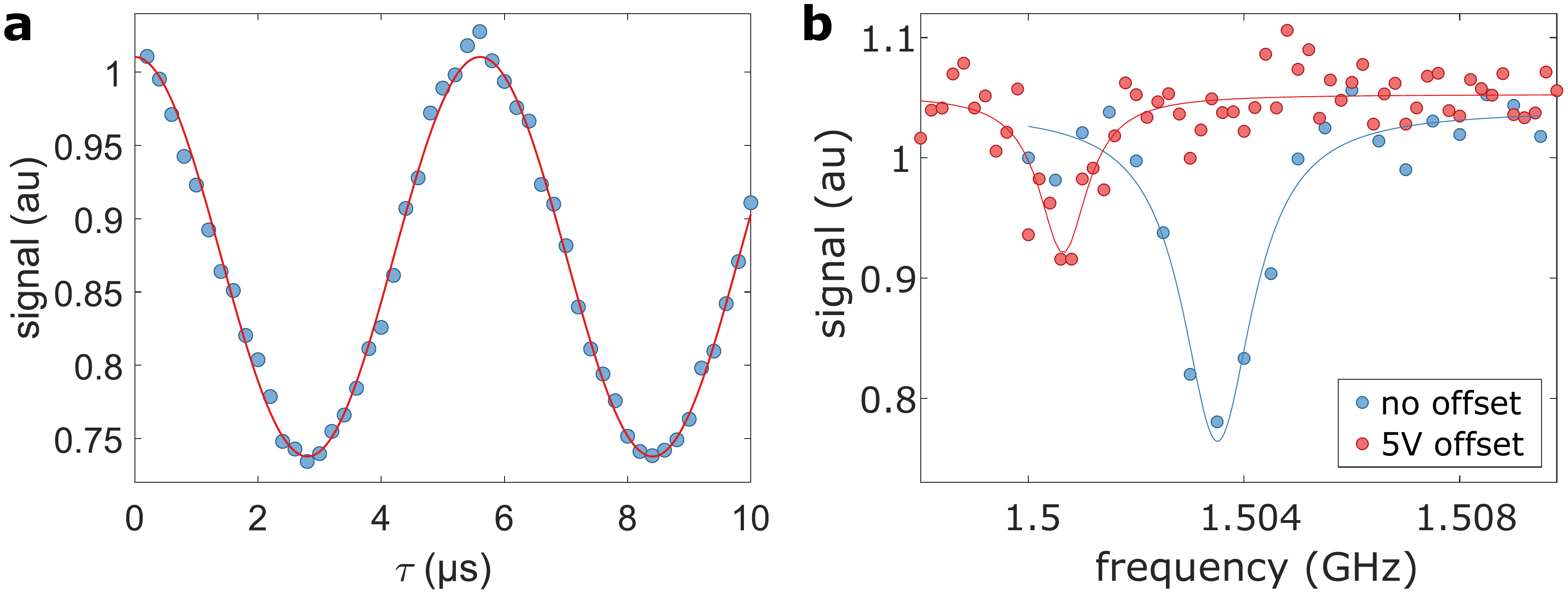}
\caption{\textbf{(a)} Rabi measurement with the signal field. With a switch the MW signal field is pulsed to perform a Rabi measurement to calibrate the signal strength. The measured signal has a Rabi frequency of $2\pi \times 178\,\mathrm{kHz}$ resulting in a magnetic field strength of $6.37\,\mathrm{\mu T}$. \textbf{(b)} Pulsed ODMR to measure the DC shift of the NV center resonance when applying a 5\,V DC voltage (red) with respect to no DC shift (blue).}
\label{fig:Rabi_DTG}
\end{figure}

Microwaves from the signal generator are controlled by an electrical switch which is operated by the AWG and allows to toggle the signal on and off in order to obtain a pulsed signal for the test measurements. To calibrate the magnetic field strength at the NV center a Rabi measurement is performed with this pulsed test signal (Figure \ref{fig:Rabi_DTG}a). From the Rabi frequency $\Omega_0$ the field strength $B_0$ is received by the relation $\Omega_0 = \gamma_\mathrm{NV} B_0$, with the gyromagnetic ratio of the NV center $\gamma_\mathrm{NV} = 2\pi \times 28.03\,\mathrm{MHz/mT}$. 

The DC shift of the resonance of the NV center is done by applying a constant current (5\,V voltage) through the copper wire. With an optically detected magnetic resonance (ODMR) measurement we determine an additional magnetic field of about one Gauss along the NV axis that results in a frequency shift of $-$3\,MHz of the $m_S = -1$ resonance (Figure \ref{fig:Rabi_DTG}b).

\subsection{Evaluation protocol} \label{sec:evaluation_protocol}
The fluorescence photons are recorded by an avalanche photo diode and its signal is directed to a time-tagged single photon counting card (TTSPC). To synchronize the photon counting card with the individual single measurements the AWG sends a trigger signal with each start of a new sequence. The TTSPC records each sequence together with the arrival time of the detected photons during the corresponding measurement with 200\,ps resolution. By post-processing of the data only photons impinging on the APD during the first 400\,ns of laser readout are selected and all other photon counts are discarded. The resulting count trace is an array of mostly zeros because the fidelity for optical readout is very low and in average about one photon in five sequences is detected. Nevertheless, the fluorescence difference of the NV center's spin states is imparted onto the fluorescence time trace as the total probability for detecting a photon is higher for $\ket{0}$ than for $\ket{\pm 1}$. Consequently, the oscillation of the spin population in the measurement protocol is imprinted onto the count trace and doing a discrete FFT reveals the beat-note frequency. 

On the (absolute valued) FFT spectrum a Lorentzian fit of the form
\begin{equation}
L(x) = L_0 \frac{\gamma^2}{(x_0-x)^2 + \gamma^2} + L_\mathrm{off}
\label{eq:Lorentzian}
\end{equation}
is performed where $L_0$ is the amplitude, $L_\mathrm{off}$ the offset, $\gamma$ the HWHM and $x_0$ the center position. The noise is determined from the square root of the variance 
\begin{equation}
\mathrm{noise} = \sqrt{\frac{1}{N-1} \sum_{i=1}^N (y(x_i)-L(x_i))^2}
\end{equation}
with the measured FFT spectrum $y(x_i)$ and the corresponding fit values $L(x_i)$. The fitting algorithm is based on a nonlinear least-squares solver. 

The uncertainty of frequency and amplitude estimation can be read from the 95\,\% error interval of the fit parameters. The uncertainty for phase estimation is calculated by error propagation of the uncertainty in the estimation of the peak position and of the noise in the real and imaginary part of the Fourier transform. Note that the peak position estimation between two data points improves with $T^{-1/2}$ corresponding only to a reduction of noise (in contrast to $T^{-3/2}$ for frequency estimation which further improves due to a reduction of the linewidth with $T^{-1}$). Therefore the scaling of the uncertainty in phase is $T^{-1/2}$. 

As the signal amplitude is estimated from the peak amplitude in the Fourier spectrum the latter has to be calibrated beforehand. This is done by measuring with a known signal. As the measurement contrast translates proportionally to the FFT amplitude a single calibration measurement is sufficient together with the results in section "Contrast in dependence on interaction time and detuning". 

\section{Measurement protocol} 
\subsection{Measurement sequence}
\label{sec:measurement_protocol}
\begin{figure}
\includegraphics[width=0.95\textwidth]{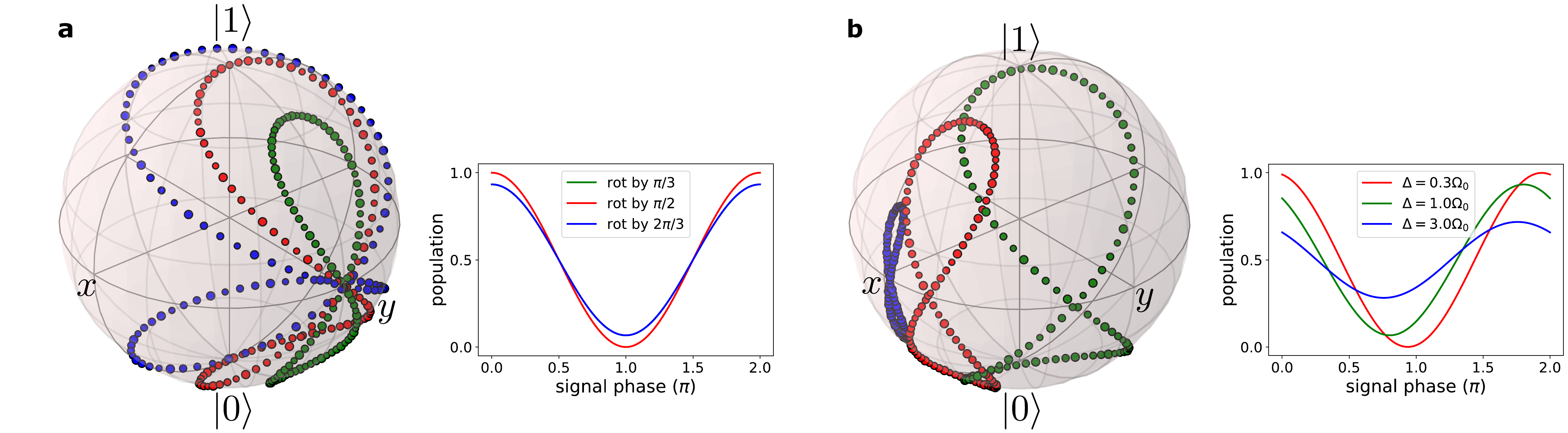}
\caption{Final states of the measurement sequence illustrated on the Bloch sphere. \textbf{(a)} Final states for phases $\phi \in [0,2\pi)$ and rotations by a resonant signal of $\Omega_\mathrm{sig}\tau =  \pi/3$, $\pi/2$ and $2\pi/3$ are shown as the green, red and blue paths, respectively. \textbf{(b)} Final states for a near-resonant signal with rotation $\Omega_\mathrm{sig}\tau = \pi/2$ and detuning $\Delta = 0.3$, $1.0$ and $3.0\,\Omega_0$ are shown as the green, red and blue paths, respectively. The graphs show the respective population of $\ket{1}$ in dependence on the signal phase $\phi$. In part \textbf{(a)} the green and blue curve give the same populations.}
\label{fig:Bloch_spheres}
\end{figure}

Our protocol for sensing high frequency AC magnetic fields is based on driving spin transitions of the NV center by the signal field. For the NV center a static magnetic field is tuned such that the energy splitting of the $\ket{0}$ and $\ket{1}$ spin states is close to the signal frequency $\nu_\mathrm{sig}$. First, the NV center is polarized into the state $\ket{0}$ during laser irradiation. To establish phase correlations between the signal and the sensor beyond the sensor's coherence time a strong control pulse of known phase is applied. In particular, we perform a resonant $\pi/2$-pulse on the NV center with phase zero. After this preparation the NV center is exposed to the signal for some interaction time $\tau$. Because the phase of the control pulse is set to zero the phase difference between the control operation and the signal is equal to the signal phase $\phi$. If we assume that the signal is on resonance with the NV spin transition and performs another $\pi/2$-rotation it is clear that for $\phi=0$ in total a $\pi$-pulse is done and the NV center's population is transferred to the $\ket{1}$ state. If the phase changes the population reduces until $\ket{0}$ is completely populated for $\phi=\pm\pi$. For $\phi = \pm\pi/2$ the rotational axis is perpendicular to the one from the control pulse and the NV center remains in an equally populated superposition state. In Figure \ref{fig:Bloch_spheres} final states for a resonant signal with different rotation angles and for a near-resonant signal are illustrated on the Bloch sphere. 

\subsection{Calculation of the final state}
\label{sec:derivation_of_pop}
In the rotating frame and after the rotating wave approximation the time evolution of a two-level system driven with a near-resonant external signal field, is described by 
\begin{equation}
\mathcal{U}_\mathrm{sig} = \cos\left( \frac{\Omega_\mathrm{sig} \tau}{2} \right) \, \mathbb{1} + 
	i \sin\left( \frac{\Omega_\mathrm{sig} \tau}{2} \right) 
	\begin{pmatrix}
	-\frac{\Delta}{\Omega_\mathrm{sig}} & \frac{\Omega_0}{\Omega_\mathrm{sig}} e^{i\phi} \\
	\frac{\Omega_0}{\Omega_\mathrm{sig}} e^{-i\phi} & \frac{\Delta}{\Omega_\mathrm{sig}}
	\end{pmatrix},
	\label{eq:time_evolution_Rabi}
\end{equation}
where the field has a constant amplitude with Rabi frequency $\Omega_0$. The duration of the interaction of the field and the system is set by the interaction time $\tau$. The signal has a phase $\phi$ and $\Delta = \omega_\mathrm{sig} - \omega_0$ is the detuning from the resonance frequency~$\omega_0$ of the two-level system giving the generalized Rabi frequency $\Omega_\mathrm{sig} = \sqrt{\Omega_0^2+\Delta^2}$. \\
In the presented measurement protocol a control $\pi/2$-pulse is applied before sensing the target field. In our settings the phase of this reference field is zero, $\phi_\mathrm{ref} = 0$. Its effect on the two-level system is
\begin{equation}
\mathcal{U}_{\pi/2} = \frac{1}{\sqrt{2}} \begin{pmatrix} 1 & i \\ i & 1 \end{pmatrix}. 
\label{eq:pi_half_evolution}
\end{equation}
The total time evolution is governed by the product of the time evolution of the two independent fields 
\begin{equation}
\mathcal{U} = \mathcal{U}_\mathrm{sig} \, \mathcal{U}_{\pi/2}.
\end{equation}
The sensor is initialized into the $\ket{0} = \begin{pmatrix} 0 \\ 1 \end{pmatrix}$ state. After evolution under $\mathcal{U}$ the amplitude of the spin $\ket{1}$ state is 
\begin{equation}
c_1 = \frac{1}{\sqrt{2}} \left[ i \cos\!\left(\frac{\Omega_\mathrm{sig}\tau}{2}\right) + 
	\sin\!\left(\frac{\Omega_\mathrm{sig}\tau}{2}\right) 
	\left( \frac{\Delta}{\Omega_\mathrm{sig}} + \frac{\Omega_0}{\Omega_\mathrm{sig}} \, 
	i e^{i\phi} \right) \right].
\end{equation}
We measure the population given by
\begin{equation}
\begin{aligned}
|c_1|^2 = \frac{1}{2} &\left[ \sin^2\!\left(\frac{\Omega_\mathrm{sig}\tau}{2}\right) \left(  
	\frac{\Delta}{\Omega_\mathrm{sig}} - \frac{\Omega_0}{\Omega_\mathrm{sig}} \sin(\phi) \right)^2 + 
	\left( \cos\!\left(\frac{\Omega_\mathrm{sig}\tau}{2}\right) + 
	\frac{\Omega_0}{\Omega_\mathrm{sig}} \sin\!\left(\frac{\Omega_\mathrm{sig}\tau}{2}\right) 
	\cos(\phi) \right)^2 \right] \\
= \frac{1}{2} &\left[ 1 - 2\frac{\Delta\Omega_0}{\Omega_\mathrm{sig}^2} 
	\sin^2\!\left(\frac{\Omega_\mathrm{sig}\tau}{2}\right) \sin(\phi) +  
	2 \frac{\Omega_0}{\Omega_\mathrm{sig}} \sin\!\left(\frac{\Omega_\mathrm{sig}\tau}{2}\right) 
	\cos\!\left(\frac{\Omega_\mathrm{sig}\tau}{2}\right) \cos(\phi) \right] \\
= \frac{1}{2} &\left[ 1 - 
	\frac{\Delta\Omega_0}{\Omega_\mathrm{sig}^2} 
	(1 - \cos(\Omega_\mathrm{sig}\tau)) \sin(\phi) + 
	\frac{\Omega_0}{\Omega_\mathrm{sig}} \sin(\Omega_\mathrm{sig}\tau) \cos(\phi) \right]
\end{aligned}
\label{eq:population_formula}
\end{equation}
as reported in the main text. One can see that for a fixed interaction time the population oscillates with the signal phase $\phi$ as the sum of a sine and cosine wave.

\subsection{Contrast in dependence on interaction time and detuning}
\label{sec:Contrast_calculations}
For periodic sampling rate the signal phase changes linearly with each measurement sequence and an oscillation of the population is obtained. To determine the contrast $C = \max_\phi(|c_1|^2) - \min_\phi(|c_1|^2)$ of the protocol, Eq. \eqref{eq:population_formula} has to be maximized and minimized as a function of the phase $\phi$. In case of zero detuning, highest and lowest population of $\ket{1}$ is received for a signal phase of 0 and $\pi$ and the contrast is $C = |\sin(\Omega_0 \tau)|$. \\
For the case of a non negligible $\Delta$ the extrema of Eq. \eqref{eq:population_formula} with respect to $\phi$ have to be found. If the derivative
\begin{equation}
\frac{d|c_1|^2}{d\phi} = \frac{1}{2} \left( -\frac{\Delta\Omega_0}{\Omega_\mathrm{sig}^2} 
	(1 - \cos(\Omega_\mathrm{sig}\tau)) \cos(\phi) + 
	\frac{\Omega_0}{\Omega_\mathrm{sig}} \sin(\Omega_\mathrm{sig}\tau) \sin(\phi) \right)
	= 0
\end{equation}
an extremum is obtained. By reordering this equation $\phi$ has to fulfill the condition 
\begin{equation}
\tan(\phi) = -\frac{\Delta}{\Omega_\mathrm{sig}} \frac{1 - \cos(\Omega_\mathrm{sig}\tau)}
	{\sin(\Omega_\mathrm{sig}\tau)} = 
	- \frac{\Delta}{\Omega_\mathrm{sig}} \tan \left(\frac{\Omega_\mathrm{sig}\tau}{2}\right).
\label{eq:phi_condition_tan}
\end{equation}
Hence an extremum is found for the phases 
\begin{equation}
\phi_k = -\arctan[\Delta/\Omega_\mathrm{sig} \tan(\Omega_\mathrm{sig}\tau/2))] + k\pi
\label{eq:phi_condition}
\end{equation}
with $k$ being an integer due to the $\pi$-periodicity of the tangent. The two phases $\phi_0$ and $\phi_1$ for example give the maximum and minimum of the population and by inserting them the contrast is calculated for any signal driving of duration $\tau$. 

\begin{figure}
\includegraphics[width=0.8\textwidth]{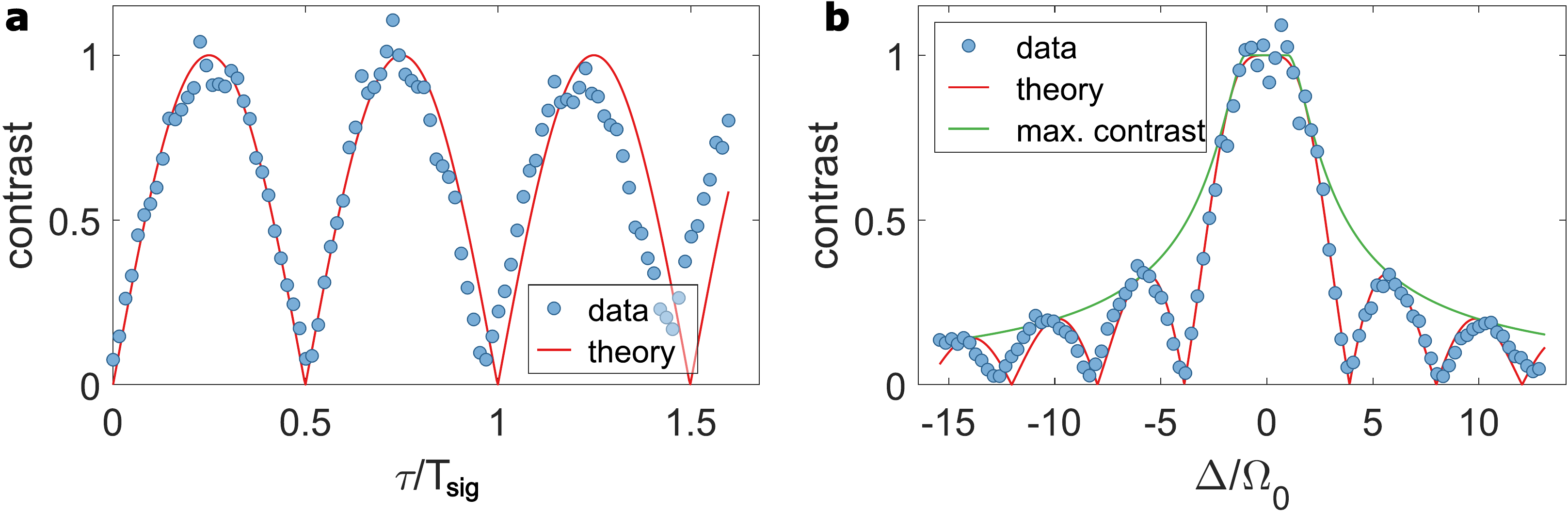}
\caption{Contrast measurements. \textbf{(a)} Contrast in dependence on interaction time with a resonant signal. On the $x$-axis the sensing time in terms of the Rabi period $T_\mathrm{sig} = 2\pi/\Omega_\mathrm{sig} \approx 2\,\mathrm{\mu s}$ is recorded. To the measurement data (blue bullets) the curve $|\sin(\Omega_0\tau)|$ is added (red line). The signal has maxima and minima for interaction lengths yielding half-integer and integer $\pi$-pulses, respectively. \textbf{(b)} Contrast with constant interaction time ($\tau=\pi/2\Omega_0 \approx 1.4\,\mathrm{\mu s}$) and in dependence on signal detuning. To the measurement data (blue bullets) a theory curve according to Eqs. (\ref{eq:population_formula},\ref{eq:phi_condition}) is added (red line). The green line shows the highest contrast achievable for optimizing $\tau$ (eq. \eqref{eq:contrast_all}). In both figures the measurement data has a slight shift from left to right with respect to the theory due to a slow drift in the transition frequency of the NV center.}
\label{fig:contrast_measurements}
\end{figure}

In Figure \ref{fig:contrast_measurements}a the peak amplitude in the FFT of the Qdyne measurement protocol is measured for a resonant signal and varying interaction time $\tau$ and in \ref{fig:contrast_measurements}b vice versa with a varying signal frequency and constant $\tau$. The amount of sampled data in the measurements is kept constant such that the peak in the Fourier transform is a measure for the contrast. The data (blue bullets) is scaled to the theoretical contrast (red line) with $C_\mathrm{meas} = a\ \mathrm{data} + \mathrm{offset}$ where the scaling factor $a$ and the offset are determined by least squares approximation. In Figure \ref{fig:contrast_measurements}b the maximum contrast that could be obtained by optimizing the sensing time is included (green line). 

For small detuning, $|\Delta| \leq \Omega_0$, full contrast can always be obtained because it is possible to perform a pulse that transfers half the population. The Rabi oscillation is described by $\Omega_0^2 / \Omega_\mathrm{sig}^2 \sin^2 (\Omega_\mathrm{sig}\tau/2)$ and has to be 1/2 to flip half the population. Hence, the interaction time has to be 
\begin{equation}
\tau = \frac{2}{\Omega_\mathrm{sig}} \arcsin \left( \frac{\Omega_\mathrm{sig}}{\sqrt{2}\Omega_0} \right).
\label{eq:tau_condition}
\end{equation}
In the case of large detuning, $|\Delta| > \Omega_0$, the amplitude of the Rabi oscillation is $\Omega_0^2/\Omega_\mathrm{sig}^2 < 0.5$. For that reason highest contrast is obtained when a $\pi$-pulse is done, i.e. when the interaction time is half the driving period, $\tau = \pi/\Omega_\mathrm{sig}$. Condition \eqref{eq:phi_condition} for the extrema yields $\phi = \pm \pi/2$ because $\tan(\Omega_\mathrm{sig}\tau / 2) = \tan(\pi/2) = \pm \infty$. For that reason if the phase difference between the control $\pi/2$-pulse and the signal is $\pm \pi/2$ highest and lowest population is received. The final population is given by $|c_1|^2 = 1/2 (1 \pm 2 \Delta\Omega_0 / \Omega_\mathrm{sig}^2)$ and the maximum contrast is
\begin{equation}  
C_\mathrm{max} = \begin{cases}
	1, & |\Delta| \leq \Omega_0 \\
	2\frac{|\Delta|\Omega_0}{\Omega_\mathrm{sig}^2}, & \mathrm{else}
	\end{cases}
\label{eq:contrast_all}
\end{equation}

\subsection{Local oscillator and beating}
In the presented protocol a local oscillator is needed with frequency close to the signal frequency. For that the resonance frequency of the sensor and hence the control field for manipulation could be used. We have identified the local oscillator as an estimate of the signal in Eq. (4) of the main text 
\begin{equation}
\nu_\mathrm{LO} = \frac{N_\mathrm{LO}}{T_L} = \frac{\mathrm{round}(\nu_\mathrm{sens}T_L)}{T_L}. 
\end{equation}
For either definition of the local oscillator the signal frequency has to be known beforehand within the sensor linewidth to assure sensitivity as well as unambiguous frequency estimation (upon sign ambiguity, see at the end of this section). All following considerations apply to any definition of the local oscillator. 

\begin{figure}
\includegraphics[width=0.8\textwidth]{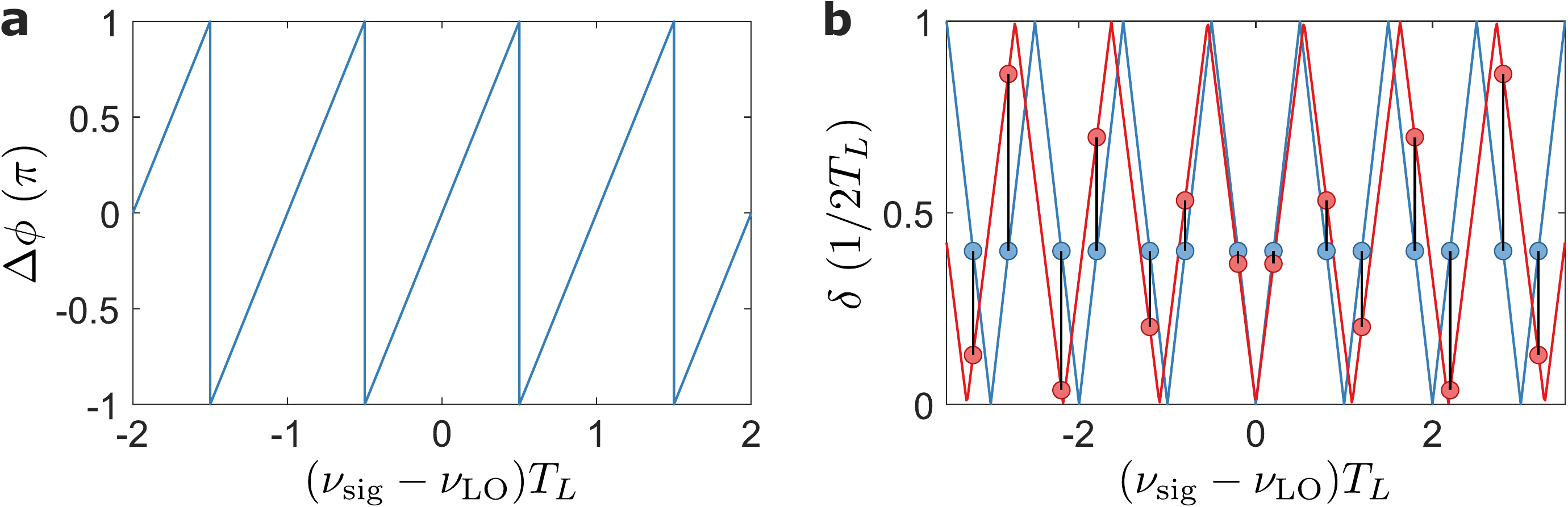}
\caption{\textbf{(a)} Phase increment $\Delta\phi$ in dependence on the signal frequency. A sawtooth curve within $[-\pi, \pi)$ and periodicity $T_L$ is obtained. \textbf{(b)} Beating as of Eq. \eqref{eq:beating_definition2} (blue). For certain signal frequencies the same beating is obtained (blue bullets). With a modified sequence length $\tilde{T}_L$ (red curve) the different frequencies can be resolved by their shift (black lines).}
\label{fig:sawtooth}
\end{figure}

We want to derive a precise formulation of the beat-note that is valid for arbitrary local oscillator and signal frequencies. The phase increment with respect to the local oscillator for each new measurement after $T_L$ is $\Delta\tilde{\phi} = 2\pi\cdot(\nu_\mathrm{sig} - \nu_\mathrm{LO}) T_L$. However, for larger values of $\nu_\mathrm{sig} - \nu_\mathrm{LO}$ ambiguities arise and we want to describe the phase increment within the interval $[-\pi,\pi)$. For that we use the sawtooth function $\mathrm{sawtooth}(x) = 2A \left( \frac{x}{p} - \lfloor \frac{1}{2} + \frac{x}{p} \rfloor \right) = 2A \left[ \left( \frac{1}{2} + \frac{x}{p} \right) \mathrm{mod}\,1 - \frac{1}{2} \right]$ with amplitude $A$ and period $p$ where $\mathrm{sawtooth}(0) = 0$. Here the floored modulo operation $r = a\ \mathrm{mod}\,n = a - n\left\lfloor \frac{a}{n} \right\rfloor$ is used. With $A = \pi$ and $p = 2\pi$ the phase increment is 
\begin{equation}
\Delta\phi = 2\pi \left[ \left( \frac{1}{2} + (\nu_\mathrm{sig} - \nu_\mathrm{LO}) T_L \right) \mathrm{mod}\,1 - \frac{1}{2} \right] = 2\pi \left[ \left( \frac{1}{2} + \delta_0 T_L \right) \mathrm{mod}\,1 - \frac{1}{2} \right]
\end{equation}
with $\delta_0 = \nu_\mathrm{sig} - \nu_\mathrm{LO}$. This function is illustrated in Figure \ref{fig:sawtooth}a. 

For the $n$-th measurement the signal phase adds up to $\phi_n = \phi_0 + n\Delta\phi$ with the initial signal phase $\phi_0$. In the end an oscillatory behavior of the population \eqref{eq:population_formula} in dependence on the measurement number $n$ with phase $\phi_n$ is observed,
\begin{equation}
|c_1^{(n)}|^2 = \frac{1}{2} \left[ 1 - 
	\frac{\Delta\Omega_0}{\Omega_\mathrm{sig}^2} (1 - \cos(\Omega_\mathrm{sig}\tau)) 
	\sin(\phi_0 + n\Delta\phi) + \frac{\Omega_0}{\Omega_\mathrm{sig}} 
	\sin(\Omega_\mathrm{sig}\tau) \cos(\phi_0 + n\Delta\phi) \right].
\label{eq:population_oscillation}
\end{equation}
The frequency of the oscillation is the beating of the local oscillator against the signal:
\begin{equation}
\delta = \frac{\Delta\phi}{2\pi T_L} = \frac{1}{T_L} \left[ \left( \frac{1}{2} + \delta_0 T_L \right) \mathrm{mod}\,1 - \frac{1}{2} \right] = \left( \frac{1}{2T_L} + \delta_0 \right) \mathrm{mod}\,\frac{1}{T_L} - \frac{1}{2T_L}
\label{eq:beating_definition}
\end{equation}
While for phase estimation the sign of $\delta$ is of importance, as pointed out in the main text, for frequency estimation the protocol does not distinguish between a positive and a negative beating. From now on we want to consider only the absolute value of the beating 
\begin{equation}
\delta = \left| \frac{\Delta\phi}{2\pi T_L} \right|	= \left| \left( \frac{1}{2T_L} + \nu_\mathrm{sig} - \nu_\mathrm{LO} \right) \mathrm{mod}\,\frac{1}{T_L} - \frac{1}{2T_L} \right|. 
\label{eq:beating_definition2}
\end{equation} 
Consequently, a sign ambiguity for estimating the signal frequency arises and $\nu_\mathrm{sig}$ cannot be distinguished from $\nu_\mathrm{sig}' = \nu_\mathrm{sig} -~2\delta_0 = 2\nu_\mathrm{LO} - \nu_\mathrm{sig}$. Assuming $0 < \nu_\mathrm{sig} - \nu_\mathrm{LO} < 1/2T_L$ the two signals $\nu_\mathrm{sig} = \nu_\mathrm{LO} + \delta$ and $\nu_\mathrm{sig}' = \nu_\mathrm{LO} - \delta$ cannot be differentiated between by measuring $\delta$. This ambiguity can be lifted by performing another measurement with a slightly modified local oscillator $\tilde{\nu}_\mathrm{LO} = \nu_\mathrm{LO} - \delta\nu$. The resulting beating is 
\begin{equation}
\tilde{\delta} = \left| \left( \frac{1}{2T_L} + \nu_\mathrm{sig} - \tilde{\nu}_\mathrm{LO} + \delta\nu \right) \mathrm{mod}\,\frac{1}{T_L} - \frac{1}{2T_L} \right| = \delta \pm \delta\nu 
\end{equation}
where the modification $\pm \delta\nu$ depends on the sign of the expression $\nu_\mathrm{sig} - \tilde{\nu}_\mathrm{LO}$. Hence, from measuring either the positive or negative shift of $\tilde{\delta}$ from $\delta$ the sign ambiguity can be resolved.

\subsection{Phase estimation} 
\begin{figure}
\includegraphics[width=0.9\textwidth]{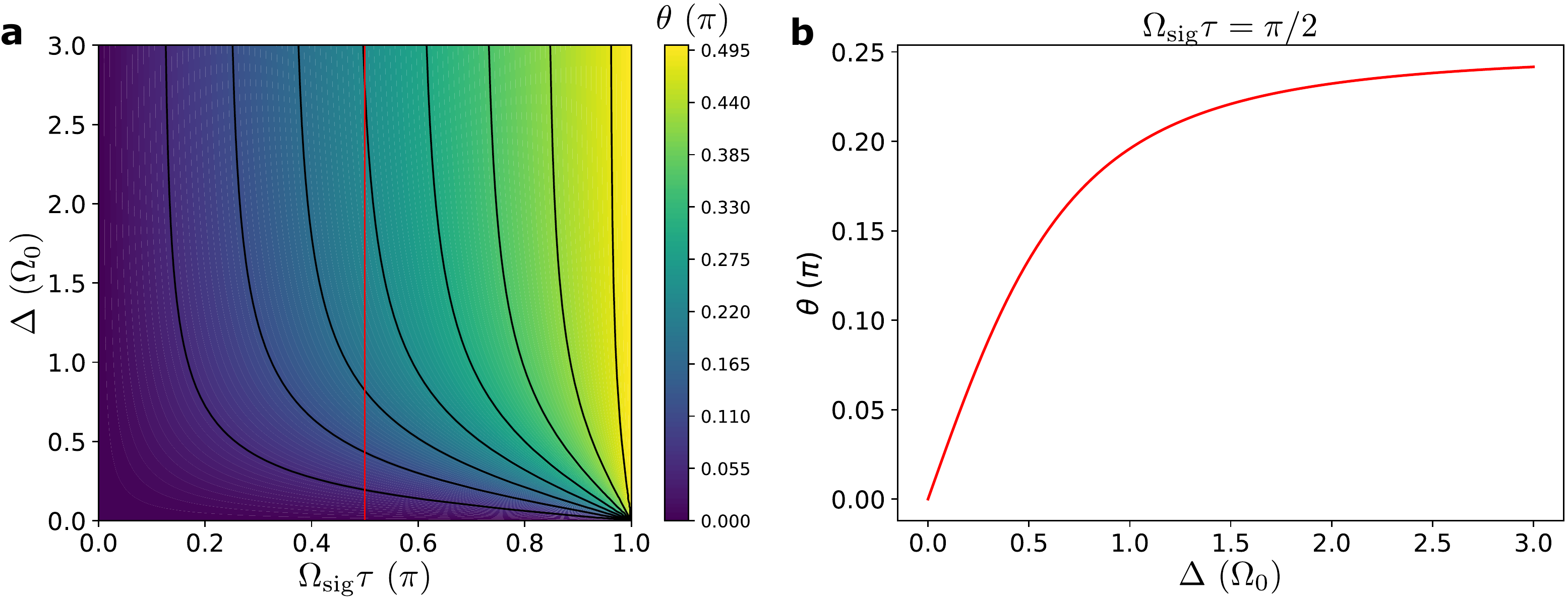}
\caption{Phase shift $\theta$. \textbf{(a)} Contour plot in dependence of the signal interaction rotation angle $\Omega_\mathrm{sig}\tau$ and the detuning $\Delta$. \textbf{(b)} Phase shift along the red line in part \textbf{(a)} for $\Omega_\mathrm{sig}\tau = \pi/2$.}
\label{fig:phase_shift}
\end{figure}

The oscillation that is measured with the Qdyne protocol is described with Eq. \eqref{eq:population_oscillation}. Importantly, it is a sum of a sine and a cosine term of same argument ($\phi_0 + n\Delta\phi$) such that the resulting oscillation has the same frequency. However, the initial phase of this oscillation might be different from the signal phase $\phi_0$. To correctly estimate the signal phase one has to take this into account. We rewrite the sine and cosine term in a single cosine expression with the identity $A \cos\phi + B \sin\phi = \mathrm{sgn}(A) \sqrt{A^2 + B^2} \, \cos(\phi + \theta)$ where the phase shift $\theta$ is given by $\theta = \arctan\left( -\frac{B}{A} \right)$. With $A = \frac{\Omega_0}{\Omega_\mathrm{sig}} \sin(\Omega_\mathrm{sig}\tau)$ and $B = - \frac{\Delta\Omega_0}{\Omega_\mathrm{sig}^2} (1 - \cos(\Omega_\mathrm{sig}\tau))$ the phase of the beat-note signal is 
\begin{equation}
\theta = \arctan\left( \frac{\Delta}{\Omega_\mathrm{sig}} \frac{1 - \cos(\Omega_\mathrm{sig}\tau)}{\sin(\Omega_\mathrm{sig}\tau)} \right) = \arctan\left( \frac{\Delta}{\Omega_\mathrm{sig}} \tan\left( \frac{\Omega_\mathrm{sig} \tau}{2} \right) \right).
\end{equation}
This shift $\theta$ can be calculated when the frequency $\nu_\mathrm{sig}$ and amplitude $B_0$ of the signal has been estimated. For small detunings $\Delta$ and rotation angles $\Omega_\mathrm{sig}\tau$ the shift is negligible, but note that in the latter case no signal is obtained. The dependence of $\theta$ on the rotation angle $\Omega_\mathrm{sig}\tau$ and the detuning $\Delta$ is shown in the contour plot in figure \ref{fig:phase_shift}a and in b as a function of $\Delta$ for $\Omega_\mathrm{sig}\tau = \pi/2$. 

\subsection{Bandwidth}
The result \eqref{eq:beating_definition2} is in accordance with the Nyquist-Shannon sampling theorem which states that the highest frequency $\delta_\mathrm{max}$ that can be resolved when sampling a signal with a constant sampling rate $1/T_L$ is given by half the sampling rate $\delta_\mathrm{max}=1/2T_L$. Signal frequencies $\nu_\mathrm{sig}^{(N)}=\nu_\mathrm{sig} + \frac{N}{T_L}$ ($N \in \mathds{Q}$) or $\nu_\mathrm{sig}'^{(N)} = \nu_\mathrm{sig} - 2\delta_0 - \frac{N}{T_L}$ (these are the sign ambiguities of $\nu_\mathrm{sig}^{(N)}$ that have the same beating with opposite sign) cannot be distinguished from $\nu_\mathrm{sig}$ because they result in the same $\delta$ (see blue bullets in Figure \ref{fig:sawtooth}b). For that reason we have assumed that the difference between the signal and local oscillator is small, $|\nu_\mathrm{sig} - \nu_\mathrm{LO}| < 1/2T_L$, such that frequencies within the interval $(\nu_\mathrm{LO}-1/2T_L, \nu_\mathrm{LO}+1/2T_L)$ can be estimated uniquely (upon sign ambiguity, see above). 

For measurements in which temporal overhead due to readout and control operations of the NV center can be neglected the sequence length is equal to the interaction time which is limited by the dephasing time, $T_L = \tau < T_2^*$. We see that the bandwidth $1/T_L > 1/T_2^*$ is always larger than the linewidth of the spin transition such that high sensitivity is ensured. If large temporal overheads ($T_L \gg \tau$) are needed (e.g. for single-shot readout) which reduce the bandwidth ($1/T_L \ll 1/T_2^*$) or to detect a wider range of frequencies to which the sensor might still be sensitive as shown in Figure \ref{fig:contrast_measurements}c of the main text, the bandwidth can be increased with a second measurement. This can be done by modifying the sequence length $\tilde{T}_L = T_L + \delta T_L$ such that the beating of the modified sequence is different for all $N$ under consideration. The beating for the $\nu_\mathrm{sig}^{(N)}$ with sequence length $\tilde{T}_L$ is
\begin{equation}
\tilde{\delta}^{(N)} = \tilde{\delta}(\nu_\mathrm{sig}^{(N)}) = \left| \left( \frac{1}{2\tilde{T}_L} + \frac{N}{T_L} + \nu_\mathrm{sig} - \nu_\mathrm{LO} \right) \mathrm{mod}\,\frac{1}{\tilde{T}_L} - \frac{1}{2\tilde{T}_L} \right| = \tilde{\delta} \pm \Delta\delta^{(N)} 
\label{eq:delta_tilde_prime}
\end{equation}
where $\tilde{\delta} = \left| \left( \frac{1}{2\tilde{T}_L} + \nu_\mathrm{sig}^{(0)} - \nu_\mathrm{LO} \right) \mathrm{mod}\,\frac{1}{\tilde{T}_L} - \frac{1}{2\tilde{T}_L} \right| $. We define the modulation of the beating as 
\begin{equation}
\Delta\delta^{(N)} = \frac{N}{T_L} \mathrm{mod} \frac{1}{\tilde{T}_L} = 
	N \frac{\tilde{T}_L}{T_L} \mathrm{mod}\, 1 \cdot 
	\frac{1}{\tilde{T}_L} = \left( N + \frac{N \delta T_L}{T_L} \right) 
	\mathrm{mod}\, 1 \cdot \frac{1}{\tilde{T}_L} 
	= \frac{N \delta T_L}{T_L} \mathrm{mod}\, 1 \cdot 
	\frac{1}{\tilde{T}_L} \stackrel{|N\delta T_L/T_L|<1}{=} 
	\frac{N \delta T_L}{T_L \tilde{T}_L}.
\label{eq:Delta_delta}
\end{equation}
However, $\Delta\delta^{(N)}$ has to fulfill a more stringent condition than that of the last equality, namely it has to ensure that $0 < \tilde{\delta}^{(N)} < 1/2T_L$. If this is not true the modulo expression of Eq. \eqref{eq:delta_tilde_prime} changes and $\Delta\delta^{(N)}$ changes accordingly. 

The same calculations can be done with $\nu_\mathrm{sig}'^{(N)}$ giving the same result
\begin{equation}
\tilde{\delta}'^{(N)} = \tilde{\delta} \pm \Delta\delta^{(N)}.
\end{equation}
Importantly, the beating frequency changes for both cases and the frequencies $\nu_\mathrm{sig}^{(N)}$ or $\nu_\mathrm{sig}'^{(N)}$ can be distinguished (sign ambiguity is not resolved) in a second measurement with modified sequence length $\tilde{T}_L$. In Figure \ref{fig:sawtooth}b the modified beating $\tilde{\delta}$ and shift $\Delta\delta^{(N)}$ are illustrated. It can also be seen for the very left and right point that here the modulo expression has changed, resulting in a different $\Delta\delta^{(N)}$ than that given in Eq. \eqref{eq:Delta_delta}.

\subsection{Unique signal estimation with large bandwidth and dynamic range} \label{sec:dynamic_range}
In the main text it is argued that with a second measurement where the interaction time $\tau$ is varied the dynamic range can be increased. At this point we want to observe that 
the three parameters local oscillator frequency, sequence length and interaction time can also be modified altogether in a second measurement. In this way unique estimation of the signal over a large bandwidth and with high dynamic range is possible. The effect of each of the modifications has to be considered as i) a constant shift but with opposite signs to resolve the sign ambiguity when modifying the local oscillator $\nu_\mathrm{LO}$, ii) a shift that increases linearly with different signal frequencies within the increased bandwidth (within the given limits) when modifying the sequence length $T_L$ and iii) a change in the beating signal amplitude when modifying the interaction time $\tau$.

\subsection{Multi-frequency signal} \label{sec:multi_freq_signal}
\begin{figure}
\includegraphics[width=0.5\textwidth]{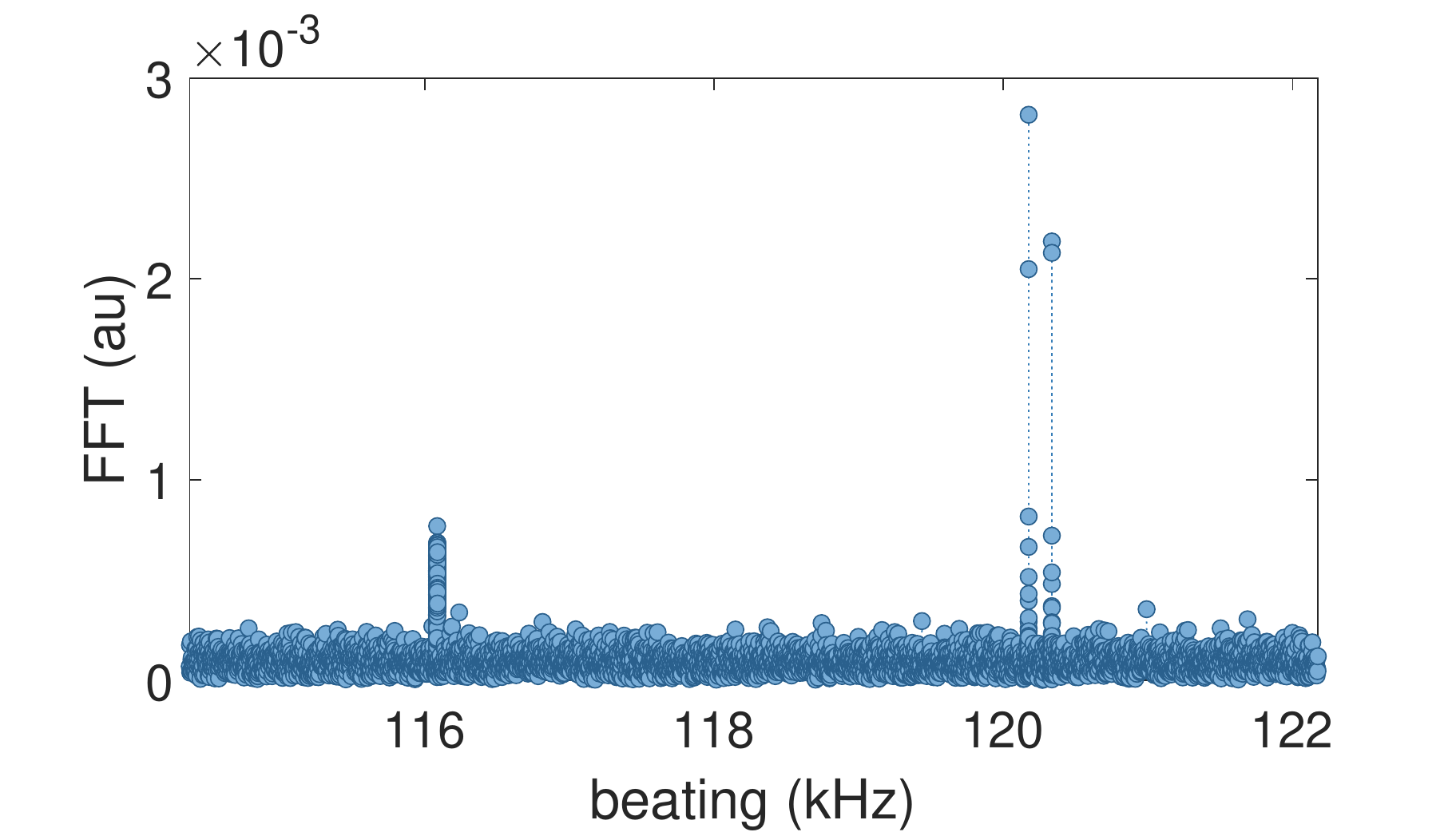}
\caption{Measurement result of a signal with three frequency components. Three MW fields with slightly different frequencies are combined to create the signal. In the Fourier transform three distinct peaks are found.}
\label{fig:three_freq_signal}
\end{figure}

With the measurement protocol it is possible to detect signals with different frequency components $\nu_i$. The action of different signal components on the Bloch sphere is that of a simultaneous rotation about different rotation axes. The result is a combined rotation of all components. With their different frequencies each one will have a different phase increment $\Delta\phi_i$ and the readout state of the sensor's spin states will change as the sum of each individual oscillation. Hence, in the Fourier transform the beat-note of each frequency component $\delta_i$ will be present. In Figure \ref{fig:three_freq_signal} the measurement result of a signal with three frequency components is shown. Each peak is resolved individually. The two peaks at the right have a frequency difference of about 160\,Hz. With the measurement protocol the two frequencies can be resolved after an integration time of $T=1/160\,\mathrm{Hz} = 6.25\,\mathrm{ms}$. The amplitude of each peak depends on the individual strength of the frequency component, its detuning and the interaction time. Within the bandwidth and the dynamic range each component can be fully reconstructed as presented for a monochromatic signal.

\end{document}